# Arbitrated quantum signature schemes without using entangled states

Xiangfu Zou[1,2] and Daowen Qiu[1,3,*]

[1]*Department of Computer Science, Zhongshan University, Guangzhou 510275, China*
[2]*Department of Mathematics and Physics, Wuyi University, Jiangmen 529020, China*
[3]*SQIG–Instituto de Telecomunicações, IST, TULisbon, Av. Rovisco Pais 1049-001, Lisbon, Portugal*

(Dated: January 25, 2010)

A digital signature is a mathematical scheme for demonstrating the authenticity of a digital message or document. For signing quantum messages, some arbitrated quantum signature schemes have being proposed. However, in the existing literature, arbitrated quantum signature schemes depend on entanglement. In this paper, we present two arbitrated quantum signature schemes without utilizing entangled states in the signing phase and the verifying phase. The first proposed scheme can preserve the merits in the existing schemes. Then, we point out, in this scheme and the prior schemes, there exists a problem that Bob can repudiate the integrality of the signatures. To conquer this problem, we construct another arbitrated quantum signature scheme without using quantum entangled states but using a public board. The new scheme has three advantages: it does not utilize entangled states while it can preserve all merits in the existing schemes; the integrality of the signature can avoid being disavowed by the receiver; and, it provides a higher efficiency in transmission and reduces the complexity of implementation. Furthermore, we present a technique such that the quantum message can keep secret to the arbitrator in a arbitrated quantum signature scheme.

PACS numbers: 03.67.Dd, 03.65.Ud

## I. Introduction

The most spectacular discovery in quantum computing to date is that quantum computer can efficiently perform some tasks which are not feasible on a classical computer. For example, Shor's quantum algorithm [1] can solve efficiently two enormously important problems: the problem of finding the prime factors of an integer and the discrete logarithm problem. This means most of the classical public key cryptography are not secure if quantum computers could be available someday. Fortunately, quantum cryptography (quantum key distribution) depends on fundamental laws of physics to provide unconditional security [2–9].

Digital signature and authentication is an essential ingredient of classical cryptography and has been employed in various applications. Similar to the classical public key cryptography, most classical digital signature schemes based on the public key cryptography can be broken by Shor's algorithm [1]. So, many researchers and scholars turn to investigate quantum signature and authentication, which is supposed to provide an alternative with unconditional security. Recently, some progress has been made on quantum signature [10–22]. In particular, an *arbitrated quantum signature* (AQS) scheme providing many merits was proposed by Zeng and Keitel [12]. This AQS scheme was further discussed in the corresponding comments [23, 24]. In such a scheme, both known and unknown quantum states could be signed, and the unconditional security is ensured by using the correlation of Greenberger-Horne-Zeilinger (GHZ) triplet states [25] and quantum one-time pads [26].

Very recently, Li et al. [13] presented an arbitrated quantum signature scheme using two-particle entangled Bell states instead of GHZ states. The scheme using Bell states can preserve the merits in the original scheme [12] while providing a higher efficiency in transmission and reducing the complexity of implementation.

We observe that the main functions of quantum entangled states (GHZ states and Bell states) in Refs. [12, 13, 24] are to assist Alice to transfer quantum states to Bob. However, Alice transfers quantum states to the arbitrator by the ciphertext encrypted with the secret key $K_A$. Similarly, Alice can transfer quantum states to Bob with a shared secret key. Considering that the prepa-ration, distribution and keeping of GHZ states and Bell states are not easy to be implemented with the present-day technologies, we construct a new arbitrated quantum signature scheme without using quantum entangled states. Furthermore, we discover that Bob can repudiate the integrality of the signature in the proposed AQS scheme and the AQS schems in Refs. [12, 13, 24]. Therefore, we give a new AQS scheme that can avoid being disavowed for the integrality of the signature by the receiver Bob.

*Electronic address: issqdw@mail.sysu.edu.cn (D.W. Qiu).

The remainder of this paper is organized as follows. First, in Section II, we briefly recall some notions and notations concerning AQS. In Section III, we give an AQS scheme similar to the schemes in Refs. [12, 13] but without using entangled states. In Section IV, we discuss the security of the scheme proposed in the previous section and point out that, in the proposed scheme and the prior schemes, there exists a problem that Bob can repudiate the integrality of the signatures. In Section V, to conquer the problem mentioned in Section IV, we give a new arbitrated quantum signature scheme without using entangled states but using a public board. In Section VI, we discuss the security of the scheme proposed in the previous section. The new scheme can conquer the problem mentioned in Section IV and preserve all merits in the foregoing schemes while providing a higher efficiency in transmission and reducing the complexity of implementation. Furthermore, we present a technique such that the quantum message can keep secret to the arbitrator. Finally, in Section VII, we make a conclusion.

In general, notation used in this paper will be explained whenever new symbols appear.

## II. Preliminaries

In this section, we briefly recall some notions and notations concerning AQS.

We use Pauli matrices $\sigma_x$ and $\sigma_z$ to denote the $X$ and $Z$ gates, respectively. Let $|P\rangle$ be a quantum message as $|P\rangle = |P_1\rangle \otimes |P_2\rangle \otimes \cdots \otimes |P_n\rangle$ with $|P_i\rangle = \alpha_i|0\rangle + \beta_i|1\rangle$.

For convenience, $E_K$ denotes the quantum one-time pads encryption, proposed by Boykin and Roychowdhury [26], according to some key $K \in \{0,1\}^*$ satisfying $|K| \geq 2n$ as follows:

$$E_K(|P\rangle) = \bigotimes_{i=1}^{n} \sigma_x^{K_{2i-1}} \sigma_z^{K_{2i}} |P_i\rangle, \quad (1)$$

where $K_j$ denotes the $j$-th bit of $K$. Similarly, $R_K$ denotes the unitary transformation

$$R_K(|P\rangle) = \bigotimes_{i=1}^{n} \sigma_x^{K_i} \sigma_z^{K_i+1} |P_i\rangle. \quad (2)$$

A secure arbitrated (quantum) signature scheme should satisfy two requirements: one is that the signature should not be forged by the attacker (including the malicious receiver) and the other is the impossibility of disavowal by the signatory and the receiver [12, 13, 24].

## III. An AQS scheme without using entangled states

From the arbitrated quantum signature schemes in Refs. [12, 24] and [13], we discover that the main functions of quantum entangled states, GHZ states and Bell states, are to assist Alice to transfer quantum states to Bob. However, Alice transfers quantum states to the arbitrator by the ciphertext encrypted with the secret key $K_A$. Similarly, Alice can transfer quantum states to Bob with a shared secret key. Considering that the preparation, distribution and keeping of GHZ states and Bell states are not easy to be implemented with the present-day technologies, we construct a new arbitrated quantum signature scheme without using entangled quantum states in the signing phase and the verifying phase.

The presented scheme also involves three participants, namely, signatory Alice, receiver Bob, and the arbitrator, and includes three phases, the initializing phase, the signing phase, and the verifying phase.

Suppose Alice need sign the quantum message $|P\rangle = |P_1\rangle \otimes |P_2\rangle \otimes \cdots \otimes |P_n\rangle$ with $|P_i\rangle = \alpha_i|0\rangle + \beta_i|1\rangle$ and has at least three copies of $|P\rangle$. For obtaining a low enough error probability in the verifying phase, we can suppose that $n$ is large enough; otherwise, we use $|P\rangle^{\otimes m}$ instead of $|P\rangle$, where $m$ is any a large enough integer.

### A. Initializing phase

*Step I1.* Alice shares the secret keys $K_{Aa}$ and $K_{AB}$ with the arbitrator and Bob, respectively, by using the *quantum key distribution* (QKD) protocols [2–4] that were proved to be unconditionally secure [5–8]. Similarly, Bob shares the secret key $K_{Ba}$ with the arbitrator.

## B. Signing phase

*Step S1.* Alice computes $|R_{Aa}\rangle = R_{K_{Aa}}(|P\rangle)$ and generates $|S_a\rangle = E_{K_{Aa}}(|P\rangle, |R_{Aa}\rangle)$.

*Step S2.* Alice computes $|R_{AB}\rangle = R_{K_{AB}}(|P\rangle)$, generates her signature $|S\rangle = E_{K_{AB}}(|R_{AB}\rangle, |S_a\rangle)$, and sends it to Bob. If they are far away from each other, they can use quantum repeaters [27, 28] and fault-tolerant quantum computation [29, 30] to ensure the signature $|S\rangle$ being transferred perfectly.

## C. Verifying phase

*Step V1.* Bob decrypts $|S\rangle$ with $K_{AB}$ and gets $|R_{AB}\rangle$ and $|S_a\rangle$.

*Step V2.* Bob generates $|Y_B\rangle = E_{K_{Ba}}(|S_a\rangle)$ and sends it to the arbitrator.

*Step V3.* The arbitrator decrypts $|Y_B\rangle$ and obtains $|S_a\rangle$. Then, he gets $|P\rangle$ and $|R_{Aa}\rangle$ from $|S_a\rangle$ with $K_{Aa}$.

*Step V4.* The arbitrator obtains $|P_a\rangle = R^{-1}_{K_{Aa}}(|R_{Aa}\rangle)$ and compares it with $|P\rangle$ using the approach in Refs. [13, 31]. If $|P_a\rangle = |P\rangle$, he sets the verification parameter $\gamma = 1$; otherwise $\gamma = 0$.

*Step V5.* The arbitrator sends the encrypted results $|Y_{aB}\rangle = E_{K'_{Ba}}(|P\rangle, \gamma)$ where the $i$th bit of $K'_{Ba}$ is the $(4n+i)$-th bit of $K_{Ba}$.

*Step V6.* Bob decrypts $|Y_{aB}\rangle$ and obtains $|P\rangle$ and $\gamma$. If $\gamma = 0$, Bob considers that the signature has been obviously forged and rejects; otherwise, he does the further verification.

*Step V7.* Bob gets $|P_B\rangle = R^{-1}_{K_{AB}}(|R_{Aa}\rangle)$ and compares it with $|P\rangle$ using the approach in Refs. [13, 31]. If $|P_B\rangle = |P\rangle$, Bob accepts the signature $|S\rangle$; otherwise, he rejects it.

## IV. Security analysis and discussion of the AQS scheme without using entangled states

A secure quantum signature scheme should satisfy three requirements [12, 13, 24]: the signature should not be forged by the attacker (including the malicious receiver); the signature should not be disavowed by the signatory; and the signature should not be disavowed by the receiver. We can show that the proposed scheme can offer security as the scheme in Refs. [12, 13]. First, we show the proposed AQS scheme without using entangled states can satisfy the first two requirements. Then, we point out that the proposed AQS scheme and the existing schemes in Refs. [12, 13] can not satisfy the third requirement.

### A. Impossibility of forgery

If the malicious receiver Bob attempts to counterfeit Alice's signature $|S\rangle = E_{K_{AB}}(|R_{AB}\rangle, |S_a\rangle)$ to his own benefit, he has to know Alice's secret key $K_{Aa}$ to construct $|S_a\rangle$. However, this is impossible due to the unconditionally secure quantum key distribution. Besides, the use of quantum one-time pad algorithm enhances the security. Thus, Bob cannot get the correct $|S_a\rangle$. Therefore, the arbitrator will discover this forgery. If the attacker Eve tries to forge Alice's signature $|S\rangle$ for her own sake, she also should know the secret keys $K_{Aa}$ and $K_{AB}$. However, the public information that he can obtain such as $|S\rangle$, $|Y_B\rangle$, and $|Y_{aB}\rangle$ betrays nothing about the secret keys $K_{Aa}$ and $K_{AB}$. Hence, the forgery for Eve is also impossible.

### B. Impossibility of disavowal by the signatory

If the signatory Alice and the receiver Bob disagree with each other, the arbitrated trusted by both of them should be required to make a judgment. Assume that Alice disavows her signature. Then the arbitrator can confirm that Alice has signed the message since the information of Alices secret key $K_{Aa}$ is involved in $|S_a\rangle$ of the signature $|S\rangle = E_{K_{AB}}(|R_{AB}\rangle, |S_a\rangle)$. Hence Alice cannot deny having signed the message.

### C. Bob can repudiate the integrality of the signature

Suppose Bob repudiates the receipt of the signature. Then the arbitrator also can confirm that Bob has received the signature $|S_a\rangle$ since he needs the assistance of the arbitrator to verify the signature. For instance, the information of his key $K_{Ba}$ is included in $|Y_B\rangle = E_{K_{Ba}}(|S_a\rangle)$. So Bob cannot disavow that he has received $|S_a\rangle$.

However, *Bob can repudiate the integrality* of the signature $|S\rangle$ because he can reject the signature in Step V7. Similarly, Bob can repudiate the integrality of the signature in the AQS schemes in Refs. [12, 13, 24].

Are there some methods to improve the AQS schemes to avoid being disavowed the integrality of the signature by Bob? We will give a new AQS scheme satisfying that the receiver Bob can not disavow the integrality of the signature.

## V. An AQS scheme unable to be disavowed by Bob

We have known that the existing AQS schemes can not avoid being disavowed for the integrality of the signature by Bob. In this section, we will present a new AQS scheme without using quantum entangled states that can avoid being disavowed for the integrality of the signature by the receiver Bob.

Note that the QKD schemes [2–4] utilize generally a public board or a classical channel that can not be blocked. Lee et al. [15] proposed an AQS scheme with a public board which can be adapted to sign classical messages. Also, we use a public board or a classical channel that can not be blocked to improve the AQS schemes to avoid being disavowed for the integrality of the signature by Bob. To avoid being disavowed by Bob, we must set the arbitrator's verifying after Bob's verifying.

The presented scheme also involves three participants, namely, signatory Alice, receiver Bob, and the arbitrator, and includes three phases, the initializing phase, the signing phase, and the verifying phase.

### A. Initializing phase

*Step I1'*. Alice shares the secret keys $K_{Aa}$ and $K_{AB}$ with the arbitrator and Bob, respectively, by using the quantum key distribution protocols [2–4] that were proved to be unconditionally secure [5–8]. Similarly, Bob shares the secret key $K_{Ba}$ with the arbitrator.

### B. Signing phase

*Step S1'*. Alice randomly chooses a number $r \in \{0,1\}^{2n}$ and computes $|P'\rangle = E_r(|P\rangle)$, and $|R_{AB}\rangle = R_{K_{AB}}(|P'\rangle)$.

*Step S2'*. Alice generates $|S_a\rangle = E_{K_{Aa}}(|P'\rangle)$.

*Step S3'*. Alice generates her signature $|S\rangle = E_{K_{AB}}(|P'\rangle, |R_{AB}\rangle, |S_a\rangle)$ and sends it to Bob. If they are far away from each other, they can use quantum repeaters [27, 28] and fault-tolerant quantum computation [29, 30] to ensure the signature $|S\rangle$ being transferred perfectly.

### C. Verifying phase

*Step V1'*. Bob decrypts $|S\rangle$ with $K_{AB}$ and gets $|P'\rangle$, $|R_{AB}\rangle$, and $|S_a\rangle$.

*Step V2'*. Bob obtains $|P'_B\rangle = R^{-1}_{K_{AB}}(|R_{AB}\rangle)$ and compares it with $|P'\rangle$ using the approach in Refs. [13, 31]. If $|P'_B\rangle = |P'\rangle$, he generates and sends $|Y_B\rangle = E_{K_{Ba}}(|P'\rangle, |S_a\rangle)$ to the arbitrator. Otherwise, he rejects the signature.

*Step V3'*. The arbitrator decrypts $|Y_B\rangle$ and obtains $|P'\rangle$ and $|S_a\rangle$ depending on the secret key $K_{Ba}$.

*Step V4'*. The arbitrator obtains $|P'_a\rangle = E^{-1}_{K_{Aa}}(|S_a\rangle)$ and compares it with $|P'\rangle$. If $|P'_a\rangle \neq |P'\rangle$, he tells Bob to reject the signature by the public board and the scheme aborts. Otherwise, he tells Alice and Bob the fact, $|P'_a\rangle = |P'\rangle$, by the public board.

*Step V5'*. Alice publishes $r$ by the public board.

*Step V6'*. Bob gets back $|P\rangle$ from $|P'\rangle$ by $r$.

## VI. Security analysis of the AQS scheme using a public board and comparison with other AQS schemes

Impossibility of forgery in the AQS scheme using a public board in Section V can be discussed as that of the AQS scheme without entangled states in Section III. Similarly, we can prove the impossibility of being disavowed by the signatory. Here, we only discuss the impossibility of being disavowed by the receiver Bob in the AQS scheme using a public board presented in Section V.

### A. Impossibility of disavowal by the receiver

It is clear that Bob must know the secret key $K_{AB}$ and $|P'\rangle = R^{-1}_{K_{AB}}(|R_{AB}\rangle)$ by Step V2'. Furthermore, Bob must have the secret key $K_{Ba}$ and $|P'\rangle = E^{-1}_{K_{AB}}(|S_a\rangle)$ by Step V3' and Step V4'. In addition, Bob can get back $|P\rangle$ from $|P'\rangle$ by Step V5' and Step V6'. By the uncondition security of the QKD and the quantum one-time pad, other people could not know both $K_{AB}$ and $K_{Ba}$. So, Bob can not disavow the receipt of the signature $|S\rangle$ and the message $|P\rangle$.

*Statement 1.* It is necessary that we only sent $|P'\rangle$ in the scheme. Bob can confirm that $|S\rangle$ is Alice's signature and get $|P\rangle$ in Step V2' if we use $|P\rangle$ instead of $|P'\rangle$ in the scheme. So, Bob need not send $|Y_B\rangle$ to the arbitrator that means Bob has a chance to disavow the receipt of the signature $|S\rangle$ and the message $|P\rangle$.

*Statement 2.* If the message $|P\rangle$ needs to keep secret to the arbitrator, we only need to modify Step V6' as "Alice publishes $r \oplus K'_{AB}$ by the public board" where the $i$th bit of $K'_{AB}$ is $(i + 6n)$-th bit of $K_{AB}$.

*Statement 3.* Similarly, the new techniques that Bob can not repudiate the integrality of the signature $|S\rangle$ and the message $|P\rangle$ can keep secret to the arbitrator, can be used to improve the prior arbitrated quantum signature schemes [12, 13].

### B. Comparing with other AQS schemes

The proposed arbitrated quantum signature scheme with a public board without using entangled states can not be disavowed by the receiver Bob while it maintains all merits of the AQS scheme using two-particle entangled Bell states in Ref. [13] and the AQS scheme using three-particle entangled GHZ states in Ref. [12]. The scheme can be adapted to both known and unknown quantum states and still provides unconditional security by employing QKD technology [2–9] and quantum one-time pads [26]. Furthermore, the AQS scheme with a public board is more efficient in the following two aspects.

One is that the total number of the transmitted qubits (bits), when $n$-qubit message is signed, is decreased as described in Table I. By Ref. [13], we know that the AQS scheme using Bell states is more efficient than that using GHZ states. So, we only need to compare it with the scheme using Bell states in Ref. [13]. Though Alice needs to publish the $2n$-bit randem string $r$, the total number of the transmitted bits and qubits is decreased significantly.

TABLE I: Comparing of the transmitted qubits quantity

| Transmission | The scheme using Bell states [13] | The scheme using a public board |
|---|---|---|
| Alice→Bob | $4n$ | $3n$ |
| Bob→The arbitrator | $4n$ | $2n$ |
| The arbitrator→Bob | $6n + 1$ | 0 |
| The arbitrator publics | 0 | a constant |
| Alice publics | 0 | $2n$ |

The other is that the complexity of implementing the scheme is reduced. Though the proposed scheme with a public board needs some local operations, it need not prepare and send Bell states and GHZ states because it does not use entangled states.

From the discussions above, we conclude that the proposed scheme with a public board achieves a higher efficiency in transmission and can be implemented easily.

### VII. Conclusions

In this paper, we have proposed two arbitrated quantum signature schemes. This two scheme can be adapted to both known and unknown quantum states and still provide unconditional security by employing QKD technology [2–9] and quantum one-time pads [26]. In the first one, we have not used quantum entangled states and proved it preserves the merits in the prior schemes [12, 13]. Furthermore, we have pointed out that there exists a problem that Bob can repudiate the integrality of the signatures in the first scheme proposed in the paper and the prior schemes [12, 13]. To conquer this problem, we have constructed a new arbitrated quantum signature scheme without using quantum entangled states but using a public board. The new scheme has three advantages. First, it does not utilize entangled states while it can preserve all merits in the existing schemes [12, 13]. Secondly, the integrality of the signature can avoid being disavowed by the receiver. In addition, it provides a higher efficiency in transmission and reduces the complexity of implementation. We have presented a technique such that the message $|P\rangle$ can keep secret to the arbitrator. Furthermore, we have pointed out that the new techniques can be used to improve on the prior arbitrated quantum signature schemes [12, 13].

### Acknowledgements

This work is supported in part by the National Natural Science Foundation (Nos. 60573006, 60873055), the Research Foundation for the Doctorial Program of Higher School of Ministry of Education (No. 20050558015), Program for New Century Excellent Talents in University (NCET) of China, and by project of SQIG at IT, funded by FCT and EU FEDER projects Quantlog POCI/MAT/55796/2004 and QSec PTDC/EIA/67661/2006, IT Project QuantTel, NoE Euro-NF, and the SQIG LAP initiative.

[1] P. W. Shor, *Proceedings of the 35th Annual Symposium on Foundations of Computer Science* (IEEE, Los Alamitos, 1994), Pages 124. P. W. Shor, SIAM Rev. **41**, 303 (1999).
[2] C. H. Bennett and G. Brassard, *Proceedings of the IEEE International Conference on Computers, Systems and Signal Processing* (IEEE, New York, 1984), p. 175.
[3] A. K. Ekert, Phys. Rev. Lett. **67**, 661 (1991).
[4] C. H. Bennett, Phys. Rev. Lett. **68**, 3121 (1992).
[5] H.-K. Lo and H. F. Chau, Science **283**, 2050 (1999).
[6] P. W. Shor and J. Preskill, Phys. Rev. Lett. **85**, 441 (2000).
[7] D. Mayers. Journal of the ACM **48**, 351 (2001).
[8] H. Inamori, N. Lütkenhaus, D. Mayers. Euro. Phys. J. D **41**, 599 (2007).
[9] M. A. Nielsen and I. L. Chuang. *Quantum Conputation and Quantum Information*. (Cambridge University Press, Cambridge, 2000).
[10] H. Barnum, C. Crépeau, D. Gottesman, A. Smith, and A. Tapp, *Proceedings of the 43rd Annual IEEE Symposium on Foundations of Computer Science* (IEEE, Los Alamitos, 2002), p. 449.
[11] M. Curty, D. J. Santos, E. Pérez, and P. García-Fernández, Phys. Rev. A **66**, 022301 (2002).
[12] G. H. Zeng and C. H. Keitel. Phys. Rev. A **65**, 042312 (2002).
[13] Q. Li, W. H. Chan, and D. Y. Long, Phys. Rev. A **79**, 054307 (2009).
[14] D. Gottesman and I. L. Chuang. arXiv: quant-ph/0105032 (2001).
[15] H. Lee, C. Hong, H. Kim, J. Lim, and H. J. Yang, Phys. Lett. A **321**, 295 (2004).
[16] X. Lü and D. G. Feng, *Proceedings of the First International Symposium on Computational and Information Science* (Springer, Berlin, 2004), p. 1054.
[17] J. Wang, Q. Zhang, and C. J. Tang, *Proceedings of the Eighth International Conference on Advanced Communication Technology* (IEEE, New York, 2006), p. 1375.
[18] X. J. Wen, Y. Liu, and Y. Sun, Z. Naturforsch. A: Phys. Sci. **62a**, 147 (2007).
[19] G. H. Zeng, M. Lee, Y. Guo, and G. Q. He, Int. J. Quantum Inf. **5**, 553 (2007).
[20] Y. G. Yang, Chin. Phys. B **17**, 415 (2008).
[21] X. Lü and D. Feng, *Proceedings of the Seventh International Conference on Advanced Communication Technology* (IEEE, Korea, 2005), p. 514 (Also, arxiv: quant-ph/04030462).
[22] Z. Cao and O. Markowitch, *Proceedings of the Sixth International Conference on Information Technology: New Generations* (IEEE, Las Vegas, 2009), p. 1574.
[23] M. Curty and N. Lütkenhaus, Phys. Rev. A **77**, 046301 (2008).
[24] G. H. Zeng, Phys. Rev. A **78**, 016301 (2008).
[25] D. M. Greenberger, M. A. Horne, and A. Zeilinger, in Bells Theorem, Quantum Theory, and Conceptions of Universe, edited by M. Kaftos (Kluwer, Dordrecht, 1989); D. M. Greenberger, M. A. Horne, A. Shimony, and A. Zeilinger, Am. J. Phys. **58**, 1131 (1990).
[26] P. O. Boykin and V. Roychowdhury, Phys. Rev. A **67**, 042317 (2003).
[27] H.-J. Briegel, W. Dür, J. I. Cirac, and P. Zoller, Phys. Rev. Lett. **81**, 5932 (1998).
[28] L.-M. Duan, M. D. Lukin, J. I. Cirac, and P. Zoller, Nature (London) **414**, 413 (2001).
[29] D. Aharonov and M. Ben-Or, *Proceedings of the 29th Annual ACM Symposium on Theory of Computing* (ACM, New York, 1997), p. 176.
[30] P. W. Shor, *Proceedings of the 37th Symposium on Foundations of Computer Science* (IEEE, Los Alamitos, 1996), p. 56.
[31] H. Buhrman, R. Cleve, J. Watrous, and R. de Wolf, Phys. Rev. Lett. **87**, 167902 (2001).